# Non-exponential Auger decay

**A.M. Ishkhanyan**[1] **and V.P. Krainov**[2]

[1]Institute for Physical Research, NAS of Armenia, 0203 Ashtarak, Armenia
[2]Moscow Institute of Physics and Technology, 141700 Dolgoprudny, Russia

**Abstract.** We discuss the possibility of non-exponential Auger decay of atoms irradiated by X-ray photons. This effect can occur at times, which are greater than the lifetime of a system under consideration. The mechanism for non-exponential depletion of an initial quasi-stationary state is the cutting of the electron energy spectrum of final continuous states at small energies. Then the Auger decay amplitude obeys $t^{-1/2}$ power-law dependence on long observation times.

## 1. Introduction

The exponential decay law of excited states of quantum systems is well known and has a rather universal character [1]. The decay is exponential, when the energy distribution of the excited state has a Lorentzian shape. The non-exponential decay, first considered by Khalfin [2], occurs in a variety of quantum systems [3-11]. For long observation times, it is caused by deviations of the energy distribution of a non-stationary state from the Lorentzian shape [3-7]. For example, the photon emitted as a result of a spontaneous decay of an excited state has a positive energy. Therefore, the energy spectrum of the emitted photons is cut at zero energy. Generally, at long observation times such cutoff results in a power-law decay of the state instead of exponential decay [3-5]. The specific law of the non-exponential decay is mostly defined by the behavior of the width of the excited state as a function of energy near the edge of the energy spectrum. If the width has a power-law dependence on the energy, then the decay also has a power-law dependence on the time. With an exponential dependence of the width on the energy, the probability for the system to stay in the initial state decreases exponentially, however, with an index of the exponent involving the time in a fractional power. Various options for such a behavior are listed in [5-6] (see also numerous references therein).

Deviations from the exponential decay law are also observed experimentally for both short [8-9] and long [10] times. The difficulty to experimentally observe these effects lies in the smallness of the decay probability. A challenge for the theory is to identify such non-stationary quantum systems, which would display larger degree of non-exponentiality, i.e.

such systems, for which the probability of staying in the initial state decreases with time as slow as possible. In the present paper we propose the Auger decay process [11] as a possible candidate for the experiment. We show that here the amplitude obeys $t^{-1/2}$ power-law dependence on the long observation times.

## 2. Non-exponential Auger decay

The Auger decay process is as follows [11]. First, the X-ray photon knocks an electron out of the K-shell of the atom (Figs. 1a,1b). For simplicity, we consider atoms of alkali elements with one valence electron. The photoionization cross section in the hydrogen-like approximation is proportional to

$$\sigma \sim Z^5 \left( \frac{\mathrm{Ry}}{\hbar\omega} \right)^{7/2}, \tag{1}$$

where $\hbar\omega$ is the energy of the photon, $\mathrm{Ry}=e^4 m_e/(2\hbar^2)$ is the Rydberg unit of energy, and Z is the effective charge acting on the emitted electron from the atomic core [1]. It is seen that the electrons are mainly emitted from the K-shell of a composite atom.

Further, the formed vacancy can be filled through a spontaneous transition of another electron from the upper shells of the formed singly charged ion (Fig. 1c). However, the probabilities of spontaneous transitions are relatively small, since they contain a factor $\alpha^3$ ($\alpha = e^2/(\hbar c) \cong 1/137$ is the fine structure constant). More probable is the nonradiative decay channel, in which the energy of such a transition is carried away through emission of a valence electron (Fig. 1d). This is the Auger decay. The probability of this process does not involve the fine structure constant. The Auger decay occurs due to the Coulomb interaction $1/|\mathbf{r}_1 - \mathbf{r}_2|$ between the valence electron 1 and the inner electron 2, which passes to the K shell from the upper shells. The matrix element of such a transition has the following form

$$V_{if} = \iint d\mathbf{r}_1 d\mathbf{r}_2 \frac{1}{|\mathbf{r}_1 - \mathbf{r}_2|} \psi_i(\mathbf{r}_1) \psi_f^*(\mathbf{r}_1) \varphi_i(\mathbf{r}_2) \varphi_f^*(\mathbf{r}_2). \tag{2}$$

Since $r_1 >> r_2$ (the valence electron is located much farther from the core than the inner electrons), then one can replace

$$\frac{1}{|\mathbf{r}_1 - \mathbf{r}_2|} \to \frac{r_2 \cos\theta_2}{r_1^2}. \tag{3}$$

Substituting (3) into (2) we get

$$V_{if} = \int d\mathbf{r}_1 \frac{\psi_i(\mathbf{r}_1) \psi_f^*(\mathbf{r}_1)}{r_1^2} \int d\mathbf{r}_2 \varphi_i(\mathbf{r}_2) \varphi_f^*(\mathbf{r}_2) r_2 \cos\theta_2. \tag{4}$$

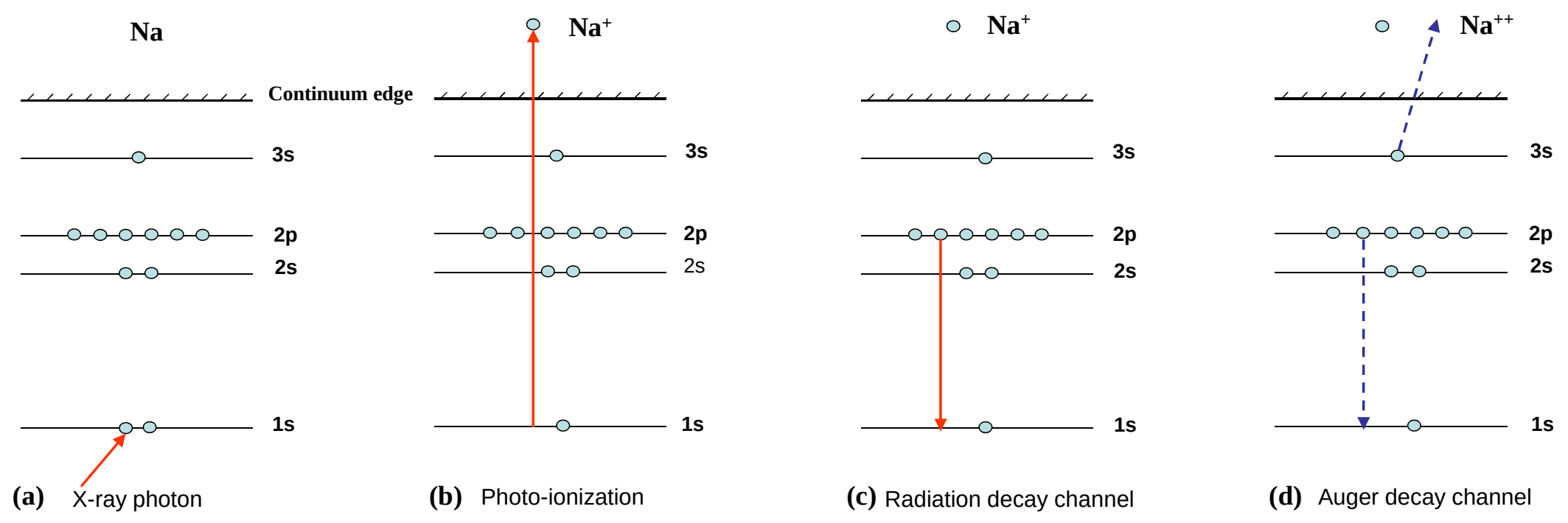

Fig.1. (a),(b) - X-ray photo-ionization, (c) - spontaneous radiation decay, (d) - Auger decay.

The second integral describes the dipole transition of the inner electron from the upper shell to the K shell. Further, as the initial state of the valence electron of an alkali atom is s-state, then, according to equation (4), the final state with the wave function $\psi_f(\mathbf{r}_1)$ is also s-state.

By $t \to \infty$, the valence electron goes to infinity with a kinetic energy $E > 0$. The probability distribution for emission with this energy is the Breit-Wigner one [1]:

$$|a(E)|^2 = \frac{1}{2\pi} \frac{\Gamma(E)}{(\Delta - E)^2 + \Gamma^2(\Delta)/4}, \tag{5}$$

where $\Delta$ is the difference between the transition energy of the inner electron and the ionization potential of the valence electron. The distribution width is $\Gamma(E) \sim |V_{if}|^2$. This distribution is normalized to unity:

$$\int_0^\infty |a(E)|^2 dE \approx 1. \tag{6}$$

As $\Gamma(E) << \Delta$, the main contribution to this integral comes from the region $E \approx \Delta$.

According to the Fock-Krylov theorem [12], the amplitude of the probability for an atom to stay in the initial state is

$$a_0(t) = \int_0^\infty |a(E)|^2 e^{-iEt} dE \tag{7}$$

or

$$a_0(t) = \frac{1}{2\pi} \int_0^\infty \frac{\Gamma(E)\exp(-iEt)\,dE}{(\Delta - E)^2 + \Gamma^2(\Delta)/4}. \tag{8}$$

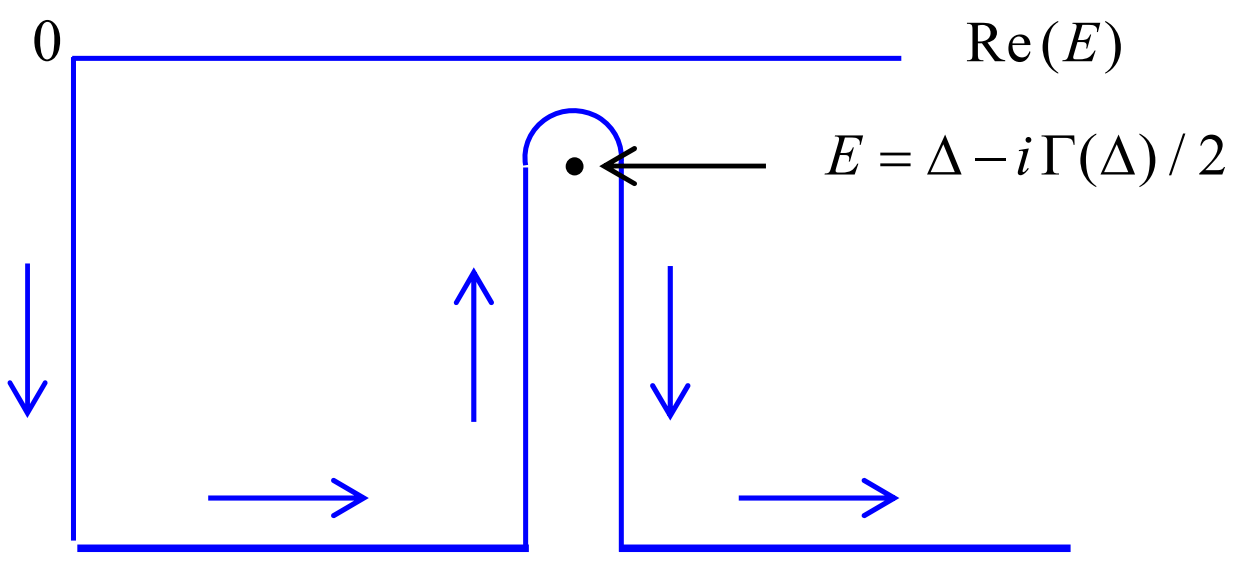

Fig. 2. Integration contour in equation (8).

The integration contour $C$ can be stretched downward up to infinity (Fig. 2). Then, at $E = \Delta - i\Gamma(\Delta)/2$, the contribution from the simple pole results in an exponential decay of the initial state:

$$a_0^{(1)}(t) = \exp\left(-i\Delta t - \Gamma(\Delta)t/2\right). \tag{9}$$

This law is obeyed at times

$$t \sim \frac{1}{\Gamma(\Delta)}. \tag{10}$$

The non-exponential decay at large times is determined by the small kinetic energies of the emitted electron, $E \to 0$. In this limit, the wave function of the electron is quasi-classical, that is $\psi_f(\mathbf{r}_1) \sim p^{-1/2}$, where $p$ is the momentum of the emitted electron [13]. Therefore, we can write

$$\Gamma(E) = \Gamma_0\sqrt{\frac{\Delta}{E}}, \quad E \to 0. \tag{11}$$

Then, by putting $E = -iz$, the integral along the vertical line on the left-hand side in Fig. 2 results in

$$a_0^{(2)}(t) = \frac{\Gamma_0 \exp(-i\pi/4)}{2\pi\Delta^{3/2}} \int_0^\infty \sqrt{\frac{1}{z}} e^{-zt} dz = \frac{\Gamma_0 \exp(-i\pi/4)}{2\pi^{1/2}\Delta^{3/2}} \frac{1}{t^{1/2}}. \tag{12}$$

Subject to the condition $a_0^{(2)}(t) > a_0^{(1)}(t)$, this non-exponential Auger-decay predominates over the exponential one, Eq. (9). This happens at times

$$t >> \frac{3}{\Gamma(\Delta)} \ln\left(\frac{\Delta}{\Gamma(\Delta)}\right). \tag{13}$$

## 3. Conclusions

Because of quite depletion of the survival probability, the observation of non-exponential decay evolution at long timescales exceeding several lifetimes poses a major experimental challenge [5]. In addition, the interaction with the environment as well as several other possible factors, including the fluctuations associated with the measurement itself, lead to persistence of the exponential regime to longer times, usually, beyond the experimental feasibility [14,15]. This is why the recent measurement of the post-exponential turnover into the power-law regime in the luminescence decay of dissolved organic molecules [10] has caused notable renewal of the interest in the subject.

As it is seen from equation (12), the non-exponential component of the amplitude in an Auger process decays in time as slow as $t^{-1/2}$. Hence, we conclude that because of slower decay rate, there is a stronger possibility to experimentally observe non-exponential behavior as compared with other discussed situations, for example, the hydrogen atom photo-ionization [16], amplification of non-Markovian decay due to bound state absorption into continuum [17], or tunnel ionization of atoms by high-intensity low-frequency laser field [6], in which the non-exponential amplitude at large times decays as $t^{-1}$. A further quest is to identify systems, which would either display larger degree of non-exponentiality or offer more robust experimental observation conditions owing to the advanced controlling techniques developed at present time for these systems. For instance, the recent unprecedented progress in experimental controlling of artificial mesoscopic semiconductor structures or ultra-cold degenerate quantum gases may offer stronger access to non-exponential regime of quantum dynamics. Apart from this particular many-particle counterpart, for which the deviations from the pure exponential law are currently poorly explored [18], further challenges include understanding of non-exponential decay in even more complex systems such as biological molecules.

## Acknowledgments

This research has been conducted within the scope of the International Associated Laboratory (CNRS-France & SCS-Armenia) IRMAS. The work has been supported by the Armenian State Committee of Science (SCS Grant No. 13RB-052). V.P. Krainov acknowledges the support from the Ministry of Education and Science of the Russian Federation (state assignment No. 3.679.2014/K).